\documentstyle[11pt,emulateapj]{article}

\received{2001 December 26}
\begin{document}

\pagestyle{myheadings}
\markboth{Tanuma, S., Yokoyama, T., Kudoh, T., \& Shibata, K. 2001,
Submitted to Astrophysical Journal}
{Magnetic Reconnection Triggered by the Parker Instability in the Galaxy}

\title{
Magnetic Reconnection Triggered by the Parker Instability in the Galaxy:\\
Two-Dimensional Numerical Magnetohydrodynamic Simulations\\
and Application to the Origin of X-Ray Gas in the Galactic Halo}

\author{\center
Syuniti Tanuma \altaffilmark{1,2,3,4,5},
Takaaki Yokoyama, \altaffilmark{4,6},
Takahiro Kudoh \altaffilmark{4,7},
\and Kazunari Shibata \altaffilmark{1,4,8}}

\altaffiltext{1}{Kwazan Observatory,
Kyoto University, Yamashina, Kyoto, 607-8471, Japan}
\altaffiltext{2}{tanuma@kwasan.kyoto-u.ac.jp}
\altaffiltext{3}{
Solar-Terrestrial Environment Laboratory, Nagoya University,
3-13 Honohara, Toyokawa, Aichi 442-8507, Japan}
\altaffiltext{4}{National Astronomical Observatory of Japan,
2-21-1 Osawa, Mitaka, Tokyo 181-8588, Japan}
\altaffiltext{5}{
Department of Astronomy, School of Science, Tokyo University
7-3-1 Hongo, Bunkyo, Tokyo 113-0033, Japan}
\altaffiltext{6}{Nobeyama Radio Observatory,
Minamimaki, Minamisaku, Nagano 384-1305, Japan
: yokoyama.t@nao.ac.jp}
\altaffiltext{7}{Department of Physics and Astronomy,
University of Western Ontario, London, Ontario N6A 3K7, Canada 
: kudoh@astro.uwo.ca}
\altaffiltext{8}{shibata@kwasan.kyoto-u.ac.jp}

\tighten
%\received{??, ???, 2001}
\accepted{??, ???, 2002}
\journalid{100}{??, ???, 2002}
\articleid{10}{20}
\paperid{AAS}
\ccc{1}
\cpright{AAS}{2002}

\begin{abstract}
We propose the Galactic flare model for the origin of the X-ray gas
in the Galactic halo.
For this purpose,
we examine the magnetic reconnection triggered by Parker instability
(magnetic buoyancy instability), 
by performing the two-dimensional resistive numerical 
magnetohydrodynamic simulations.
As a result of numerical simulations,
the system evolves as following phases:
Parker instability occurs in the Galactic disk.
In the nonlinear phase of Parker instability,
the magnetic loop inflates from the Galactic disk into the Galactic halo,
and collides with the anti-parallel magnetic field,
so that the current sheets are created in the Galactic halo.
The tearing instability occurs, and creates the plasmoids (magnetic islands).
Just after the plasmoid ejection,
further current-sheet thinning occurs in the sheet,
and the anomalous resistivity sets in. 
Petschek reconnection starts,
and heats the gas quickly in the Galactic halo.
It also creates the slow and fast shock regions in the Galactic halo.
The magnetic field ($B\sim 3\ \mu$G), for example, can heat the gas
($n\sim 10^{-3}$ cm$^{-3}$) to temperature of $\sim 10^6$ K
via the reconnection in the Galactic halo.
The gas is accelerated to Alfv\'en velocity ($\sim 300$ km s$^{-1}$).
Such high velocity jets are the evidence of the Galactic flare model
we present in this paper,
if the Doppler shift of the bipolar jet is detected in the Galactic halo.
\\
%\keywords{
%MHD --- numerical simulation ---
%ISM: kinematics and dynamics --- ISM: magnetic fields ---
%Galaxy: structure --- X-rays: ISM}

\end{abstract}
%\keywords{ISM: magnetic fields --- instabilities}
\keywords{Galaxy: halo --- ISM: magnetic fields --- instabilities --- magnetohydrodynamics}

\section{INTRODUCTION}

The X-rays from hot gas are observed in the Galactic halo.
Its luminosity and temperature are
$L_{\rm X}\sim 7\times 10^{39}$ erg s$^{-1}$ and $T\sim 10^6$ K
(\cite{pie98}).
The volume and thermal energy are estimated to be 
$E\sim 10^{55}$ erg and $V\sim 10^{68}$ cm$^3$.
To explain such hot gas in the Galactic halo,
the ``Galactic fountains'' model has been proposed
(i.e., supernova remnants and stellar winds heat the gas;
\cite{bre80}; \cite{nor89}; \cite{sha91}; \cite{shu96};
\cite{avi00}; \cite{sla00}).
The energy source, however, may not be explained fully by this model
because the evidence is not observed adequately
(see also Birk, Lesch, \& Neukirch 1998).
In this paper, we propose another mechanism.
It is the Galactic flare model, i.e., the magnetic heating in the Galactic halo.

Parker (1992) pointed out the importance of magnetic reconnection 
for the heating of Galactic plasmas.
Whenever the magnetic flux collides with another flux, 
which is not exactly parallel, the current density increases,
and a strong dissipation sets in (e.g., via anomalous resistivity)
to trigger fast reconnection (e.g., \cite{uga86}).
The magnetic reconnection is observed in solar flares 
by the X-ray satellites {\it Yohkoh} (\cite{mas94}; \cite{shi96}; \cite{tsu96})
and {\it SoHO} (\cite{yok01}).
In the solar flare, the reconnection heats the plasma from a temperature 
of $\sim$ several $\times 10^6$ K to $\sim$ several $\times 10^{7-8}$ K,
and accelerates it to Alfv\'en velocity ($\sim 10^{2-3}$ km s$^{-1}$)
(e.g., \cite{shi96}).
The reconnection would occur also in the Galaxy 
(Tanuma et al.\ 1999a, 1999b, 2001a, 1999b),
which may be called ``Galactic flare'' (\cite{stu80}; \cite{kah93}).
Total magnetic energy is
$E_{\rm mag}\sim$
$(\langle B\rangle_{\rm obs}^2/8\pi)$ $V_{\rm G}$
$\sim 10^{54.4}$ erg at least,
where $\langle B\rangle_{\rm obs}$($\sim 3\ \mu$G)
is the mean observed field strength (see \cite{bec96}; \cite{val97}),
and $V_{\rm G}$ ($\sim 10^{67}$ cm$^{3}$) is the volume of the Galaxy.
The rotational energy of the Galaxy ($\sim 10^{58.9}$ erg)
and kinetic energy of the interstellar gas ($\sim 10^{58.2}$ erg) 
are its origin (e.g., \cite{par71}; \cite{stu80}; \cite{tan99a}; \cite{tan00}).

The steady reconnection mechanisms were proposed (see \cite{pri00}).
In Sweet(1958)-Parker(1957) type reconnection,
the diffusion region is so long as to occupy whole current system.
It can not be applied to the solar flare phenomena,
because the reconnection rate of this model is too small
($\sim R_{\rm m}^{-1/2}$) in the solar corona,
where $R_{\rm m}$ is the magnetic Reynolds number ($\sim 10^{12}$).
On the other hand, in Petschek(1964) type reconnection,
the diffusion region is localized near an X-point,
and standing slow shocks occupy whole current systems.
In this case, the energy conversion via slow shocks
is much larger than Ohmic heating.
This can hence be applicable to the solar flare phenomena,
because the reconnection rate of this model is $\sim 0.1-0.01$.
This is called ``fast reconnection''.
A basic problem of Petschek model is as follows:
Petschek reconnection occurs, 
if the anomalous resistivity sets in the current sheet
(e.g., \cite{uga86}; \cite{yok94}; \cite{tan00}; 
Tanuma et al.\ 1999a, 2001a).
The anomalous resistivity set in,
when the current-sheet thickness becomes comparable 
with ion Lamor radius or ion inertial radius.
It is, however, not fully known how the current sheet becomes thin,
because the typical size of solar flare ($10^{9-11}$ cm) is 
much larger than these radii ($10^{2-3}$ cm).
This situation is similar to that of the Galaxy ($R_{\rm m}>10^{15}$),
where typical size of magnetic field ($>10^{19}$ cm)
is much larger than the ion Lamor radius ($\sim 10^7$ cm).
To solve these problems, we proposed the current-sheet thinning 
via the ``fractal tearing instability'' 
(\cite{tan00}; \cite{shi01}; \cite{tan01a}).

Many two-dimensional (2D) magnetohydrodynamic (MHD) numerical simulations 
have been carried out for the magnetic reconnection in the solar atmosphere
(\cite{mag97}; \cite{ods97}),
and in the Galactic halo (Zimmer, Lesch, \& Birk 1997; \cite{bir98}),
by assuming the current sheet at the initial condition
(see also \cite{nit01}).
Recently, Tanuma et al.\ (1999a, 1999b, 2001a) examined
the magnetic reconnection triggered by a supernova-shock 
by performing the 2D MHD simulations with a high spatial resolution,
and proposed that it can generate X-ray gas
in the Galactic disk (e.g., \cite{ebi01}).
They found that the tearing instability (\cite{fur63})
occurs in the current sheet long after the passage of a shock wave.
Petschek reconnection occurs after further current sheet thinning
via secondary tearing instability.
In the present model, Parker(1966) instability creates the current sheet 
by itself and trigger the magnetic reconnection 
(\cite{tan99b}; \cite{tan00}; see also Shibata, Nozawa, \& Matsumoto 1992;
Yokoyama \& Shibata 1996, 1997).

Recently, we examied three-dimensional (3D) MHD simulations of 
the magnetic reconnection with a low spatial resolution.
Petschek reconnection occurs after the current sheet thinning
by the tearing instability in both 2D (\cite{tan99b})
and 3D models (\cite{tan00}; \cite{tan01b}),
because we can not resolve the secondary tearing instability
when we assume a rough grid
(The similarities between 2D and 3D models are consistent with
Ugai \& Shimizu 1996).
3D effect such as Rayleigh-Taylor instability, however, appears 
when reconnection jet collides with high pressure gas and magnetic loop
much after the onset of Petschek reconnection.
Tanuma et al.\ (2002) applied the results to 
the creation of helical magnetic field
and confinement of high energy particles in the solar flare.
We study the basic physics of magnetic reconnection which are
common between 2D and 3D models,
although 2D model examined in this paper is a toy model of limited 
in 2D dimension.
In this paper, however, 
we examine the 2D model under a higher spatial resolution than 
the 3D model which we are able to do.

Parker instability is the undular mode
($\mbox{\boldmath$k$}\parallel \mbox{\boldmath$B$}$)
of magnetic buoyancy instability (\cite{par66}),
which occurs if a gas layer in a gravitational field
is supported by the horizontal magnetic fields.
Suppose that the magnetic field lines are disturbed
and begin to undulate.
The gas in the loop top slides down along the field lines,
so that loop rises further, and the instability sets in.
Parker instability is suggested to influence
the motion of clouds, H{\small II} regions,
and OB associations (\cite{tos74}), and the distribution of clouds
(\cite{mou74}; \cite{vrb77}; \cite{bli80}; \cite{elm82});
for example, Perseus hump (\cite{sof74}), 
Perseus arm (Appenzeller 1971, 1974), Barnard loop (\cite{mou74}),
and Sofue-Handa(1984) lobe.

The linear analysis of Parker instability were made by many researchers 
(\cite{shu74}; \cite{hor88}; \cite{han92}; \cite{fog94}; 
\cite{cho97}; \cite{kam97}).
The 2D MHD simulations were performed for solar flares
(\cite{kai90}; \cite{noz92}; \cite{shi92}; Yokoyama \& Shibata 1996),
and the interstellar medium
(\cite{bas97}; \cite{mat98}; \cite{san00}; \cite{ste01}).
The three-dimensional (3D) simulations of Parker instability
of horizontal magnetic field in solar atmosphere and Galaxy
(\cite{mat93}; \cite{kim01}; \cite{han02}),
and the twisted flux tube in the solar atmosphere
(\cite{mat98}; \cite{abb00}; \cite{fan01}; \cite{mag01}),
the Galaxy (\cite{han00}; \cite{fra02}), 
and accretion disks (\cite{zie01}) are also performed.
Shibata et al.\ (1989, 1992) and Yokoyama \& Shibata (1994, 1996, 1997)
examined the magnetic reconnection triggered by Parker instability
in the solar corona.
Recently, Hanasz et al.\ (2002) examined the 3D model
of the magnetic reconnection in magnetic loop created by
Parker instability with Coriolis force in the interstellar medium.
In the present paper, we extend Shibata et al.\ (1989, 1992) 
and Yokoyama \& Shibata (1994, 1996, 1997)'s solar flare model to the Galaxy: 
The Galactic flare as the origin of the X-ray gas in the Galactic halo.

In this paper,
we propose a possible origin of X-ray gas in the Galactic halo.
In the next section, we describe the simulation method.
In sections 3 and 4, 
we describe the results of numerical simulations, and discuss it.
In the last section, we summarize this paper.

\section{NUMERICAL SIMULATIONS}

\subsection{The Situation of the Problem}

Figure \ref{picture} shows our schematic scenario.
Figure \ref{picture}{\it a} displays the initial condition.
Parker instability occurs in the Galactic disk (Fig. \ref{picture}{\it b}; 
we call this situation phase I and II in this paper).
The magnetic reconnection occurs and heats the gas 
(Fig. \ref{picture}{\it c}; phase III) .
The heated gas is confined by the magnetic field (Fig. \ref{picture}{\it d}).

\subsection{Two-Dimensional Resistive MHD Basic Equations}

The resistive MHD basic equations are written as follows:
\begin{eqnarray}
{\partial\rho\over\partial t}+\nabla\cdot (\rho\mbox{\boldmath$v$})&=&0,\\
\rho{\partial\mbox{\boldmath$v$}\over\partial t}
+\rho (\mbox{\boldmath$v$}\cdot\nabla)\mbox{\boldmath$v$}
+\nabla p_g
&=& {1\over c} \mbox{\boldmath$J$}\times\mbox{\boldmath$B$}
+\rho\mbox{\boldmath$g$},\\
{\partial\mbox{\boldmath$B$}\over\partial t}
-\nabla\times (\mbox{\boldmath$v$}\times\mbox{\boldmath$B$})
&=& -c\nabla\times(\eta\mbox{\boldmath$J$}),\\
{\partial e\over\partial t}
+\nabla\cdot\left[ (e+p_g)\mbox{\boldmath$v$}\right]
&=& \eta |\mbox{\boldmath$J$}|^2
+\mbox{\boldmath$v$}\cdot\nabla p_g
\end{eqnarray}
where $\rho$, $\mbox{\boldmath$v$}$, $\mbox{\boldmath$B$}$,
$\eta$, $e$, $\mbox{\boldmath$J$}$, and $\mbox{\boldmath$g$}$
are mass density, velocity, magnetic field,
electric resistivity, internal energy,
current density ($=c\nabla\times\mbox{\boldmath$B$}/4\pi$),
and gravity, respectively.
We use the equation of state for the ideal gas,
i.e., $p_g=(\gamma-1)e$,
where $\gamma$ is the specific heat ratio (=5/3).
We take Cartesian coordinates ($x,z$),
where $x$ and $z$ axes are in horizontal and vertical directions,
respectively.
The velocity, magnetic field, and gravitational acceleration are 
$\mbox{\boldmath$v$}=(v_x, v_z)$, $\mbox{\boldmath$B$}=(B_x, B_z)$,
and $\mbox{\boldmath$g$}=[0,-g\tanh(z/0.001)]$, 
where $g$ is constant.

\subsection{The Normalization}

We normalize the velocity, length, and time
by the sound velocity ($C_s$), gravitational scale height ($H$),
and $H/C_s$, respectively.
The units of normalizations are
\begin{eqnarray}
C_s
  &\sim& 10 T_4^{1/2}\ \rm km\ s^{-1},
\label{simpiCs}\\
H % &\sim& {RT\over\rho}
  &\sim& {C_s^2\over g}\sim 100 T_4^{1/2}\ \rm pc,
\label{simpiH}\\
\tau
  &\equiv& {H\over C_s}\sim 10^7\ \rm yr,
\end{eqnarray}
where $T_4$ is the temperature in unit of $10^4$ K,
and $g$ is the acceleration of the gravity (\cite{tan99b}; \cite{tan00}).
The units of density, gas pressure, magnetic field strength, 
current density, and resistivity are 
$\rho_0\sim 1\times 10^{-25}$ g cm$^{-3}$,
$\rho_0 C_s^2\sim 4\times 10^{-13}$ erg cm$^{-3}$,
$(\rho_0 C_s^2/8\pi)^{1/2}\sim 3.2\ \mu$G,
$(\rho_0 C_s^2/8\pi)^{1/2}/H$,
and $c^2HC_s/4\pi$, respectively.
In this paper, we solve the non-dimensional basic equations,
and describe the results of numerical simulations
by non-dimensional values.

\subsection{The Anomalous Resistivity Model}

We assume the anomalous resistivity model as follows:
\begin{equation}
\eta=\left\{\begin{array}{ll}
0                   & \mbox{if}\ v_d\leq v_c\\
\alpha(v_d/v_c-1)^2 & \mbox{if}\ v_d  >  v_c
            \end{array}\right.
\label{eta}
\end{equation}
(e.g., Ugai 1986, 1992; \cite{shi92}; \cite{yok97}; 
\cite{uga01}; Tanuma et al.\ 1999b; \cite{tan00}),
where $v_d$($\equiv |J|/\rho$) and $v_c$
are the relative ion-electron drift velocity
and threshold above which anomalous resistivity sets in.
The parameters are $\alpha=0.01$ and $v_c=400$.

\subsection{The Initial Condition of Typical Model}

For the initial condition, we assume that 
the cool, dense Galactic disk is between the hot, rarefied Galactic halo 
(Fig. \ref{picture}{\it a}; 
\cite{hor88}; \cite{mat88}; \cite{tan99b}; \cite{tan00}).
Tables \ref{parameter} and \ref{normalization} show 
the variables and parameters.
The temperature is 
\begin{equation}
T(x,z)=T_0
+0.5
\left[\tanh\left({z-z_c\over 0.5}\right)+1.0\right](T_c-T_0).
\end{equation}
The parameters are $T_0=1$, $T_c=25$, and $z_c=6$.
The basic physics does not much depend on $T_c/T_0$ 
(\cite{mat88}; \cite{yok96}).
The magnetic fields are in the Galactic disk ($|z|<z_d=2.5$)
and in the Galactic halo ($|z|>z_h=5$).
The initial magnetic field is horizontal ($B_z(x,z)=0$) everywhere,
so that the ratio of gas to magnetic pressure is 
$\beta_0=8\pi p_g(x,z)/B_x(x,z)^2=0.2$.
The $\beta(z)$ is 
\begin{eqnarray}
{1\over\beta(z)}
&=&{1\over 2\beta_0}
 \left[1-\tanh\left({z-z_d\over 0.5}\right)\right]\nonumber\\
&+&{1\over 2\beta_0}
 \left[1+\tanh\left({z-z_h\over 0.5}\right)\right].
\end{eqnarray}
The value ($\beta_0=0.2$) is attained in the Galactic disk, 
if the magnetic field of $B\sim$ $5^{1/2}\langle B\rangle_{\rm obs}$ 
$\sim 7\ \mu$G,
where $\langle B\rangle_{\rm obs}$ is 
the mean observed field strength ($\sim 3\ \mu$G) in the Galactic disk
(see \cite{bec96}; \cite{val97}).
The magnetized gas is under the MHD equilibrium;
\begin{equation}
{d\over dz}\left[p_g(z)+{B_x(z)^2\over 8\pi}\right]+\rho(z)g=0.
\label{pbz}
\end{equation}
The variables are $\rho_0=1$, $p_{g0}=0.6$, and $B_{x0}\simeq 8.68$
at the equatorial plane.
The sound and Alfv\'en velocity are
$C_s\equiv (\gamma p_{g0}/\rho_0)^{1/2}=1.0$ and
$v_{\rm A}^{\rm init}=B_{x0}/(4\pi\rho_0)^{1/2}\simeq 2.45$
in the Galactic disk.
The Galactic disk is unstable to Parker instability.

We use the $N_x=403$ grids and $N_z=604$ grids
in the horizontal and vertical directions, respectively.
The intervals of grids are uniform ($\triangle x=0.30$) in $x$-axis
and nonuniform ($\triangle z\geq 0.075$) in $z$-axis.
We assume the top ($z=+34.5$) and bottom ($z=-34.5$) surfaces
are free boundaries, and the right ($x=+30.0$) and left ($x=-30.0$)
ones are periodic ones.
We use the 2-steps modified Lax-Wendroff method.
We put random perturbations ($<0.05C_s$)
on vertical velocities [$v_z(x,z)$] in the Galactic disk.

\bigskip

We neglect radiation and heat conduction.
The cooling times due to the conduction and radiation
for the cool ($T\sim 10^4$ K), dense ($n\sim 0.1$ cm$^{-3}$) gas
in the Galactic disk are
\begin{eqnarray}
\tau_{\rm cond}
&\sim& {nkT\lambda^2\over\kappa_0 T^{7/2}}\\
&\sim& 10^{13}\left({n\over 0.1\ {\rm cm}^{-3}}\right)\nonumber\\
& &           \left({\lambda\over 100\ {\rm pc}}\right)^2
              \left({T\over 10^4\ {\rm K}}\right)^{-5/2}\ {\rm yr},\\
\tau_{\rm rad}
&\sim& {nkT\over n^2\Lambda(T)}\\
&\sim& 10^3\left({T\over 10^4\ {\rm K}}\right)
           \left({n\over 0.1\ {\rm cm}^{-3}}\right)^{-1}\nonumber\\
& & \left[{\Lambda(10^4\ {\rm K})
     \over 10^{-21}\ {\rm erg\ cm^3\ s^{-1}}}\right]^{-1}\ {\rm yr},
\end{eqnarray}
respectively,
where $\Lambda(T)$ is the cooling function (\cite{spi78}),
and $\kappa_0$ is constant ($=10^{-6}$ erg s$^{-1}$ cm$^{-1}$ K$^{-1}$).
For the hot ($T\sim 10^6$ K), rarefied ($n\sim 10^{-3}$ cm$^{-3}$) gas 
in the Galactic halo, however, they become 
\begin{eqnarray}
\tau_{\rm cond}
&\sim& 10^9 \left({n\over 10^{-3}\ {\rm cm}^{-3}}\right)\nonumber\\
& & \left({\lambda_{\rm eff}\over 3\ {\rm kpc}}\right)^2
    \left({T\over 10^6\ {\rm K}}\right)^{-5/2}\ {\rm yr},\\
\label{cond2}
\tau_{\rm rad}
&\sim& 10^9 \left({T\over 10^6\ {\rm K}}\right)
            \left({n\over 10^{-3}\ {\rm cm}^{-3}}\right)^{-1}\nonumber\\
& & \left[{\Lambda(10^6\ {\rm K})
   \over 10^{-23}\ {\rm erg\ cm^3\ s^{-1}}}\right]^{-1}\ {\rm yr},
\end{eqnarray}
respectively, where 
$\lambda_{\rm eff}$ is the effective length of helical magnetic loop.
They are one order of magnitude longer than the typical time scale
($\sim 10^8$ yr; see next section)
of the physical process examined in this paper,
so that the cooling mechanisms can be neglected,
once they are heated to X-ray gas in the Galactic halo.
The basic properties of the magnetic reconnection
such as reconnection rate and energy release rate are not much
affected by the heat conduction (\cite{yok97}).

\section{THE RESULTS OF NUMERICAL SIMULATIONS}
\subsection{The typical Model (Model A1)}

\bigskip
{\bf Phase I: The Linear Phase of Parker Instability ($t<40$)}

Figures \ref{te}-\ref{j} show the time evolution of the system.
The axes are in the unit of $\sim 10H\sim 1$ kpc.
The magnetic field lines starts to bent across the plane.
Figures \ref{te}{\it a}, \ref{de}{\it a}, and \ref{j}{\it a} shows
a characteristic feature of the odd-mode (glide-reflection mode),
which grows earlier than even-mode (mirror-symmetry mode) 
(Horiuchi et al.\ 1988; Matsumoto et al.\ 1988;
see also \cite{tan99b}; \cite{tan00}).
The magnetic field inflates toward the Galactic halo
by the magnetic buoyancy force.
The gas slides down along the magnetic field lines.

\bigskip
{\bf Phase II:
The Nonlinear Phase of Parker Instability ($t\sim 40-60$)}

The system enters the nonlinear phase at $t\sim 40$.
In the valleys of the waving field,
the vertical dense spurs are formed almost perpendicular to 
the Galactic plane (Fig. \ref{de}{\it a}).
The dense regions are also created in the valleys.
Figure \ref{vx} shows the time variations of the drift velocity and velocity.
The velocity increases to Alfv\'en velocity ($\sim 2.5-3.0$) in this phase.

Figure \ref{energy} shows the time variation of the various energies.
The magnetic, thermal, kinetic, gravitational, and total energies are defined by
\begin{eqnarray}
E_{\rm mag}&=&\int\int 
{B_x(x,z)^2+B_z(x,z)^2\over 8\pi} {\rm d}x{\rm d}z,\\
E_{\rm th} &=&\int\int {1\over \gamma-1} p_g(x,z) {\rm d}x{\rm d}z,\\
E_{\rm kin}&=&\int\int
{1\over 2} \rho(x,z)\left(v_x(x,z)^2+v_z(x,z)^2\right){\rm d}x{\rm d}z,\\
E_{\rm g}  &=&\int\int \rho(x,z) g |z| {\rm d}x{\rm d}z,\\
E_{\rm tot}&=& E_{\rm mag} + E_{\rm th} + E_{\rm kin} + E_{\rm g},
\end{eqnarray}
respectively.
They are integrated in whole simulation region,
and corrected by the energy flux at the top and bottom boundaries.
The gravitational and magnetic energies 
are converted to the thermal and kinetic ones.

The top of inflating loops collides with the anti-parallel magnetic field
in the Galactic halo (Figs. \ref{te}{\it b} and \ref{de}{\it b}),
so that the current sheets are created at $t\sim 60$ 
(Fig. \ref{j}{\it b}).

\bigskip
{\bf Phase III: The Magnetic Reconnection Triggered by 
Tearing Instability ($t>60$)}

The drift velocity increases steeply at $t\sim 60$ (Fig. \ref{vx}{\it a}).
The tearing instability occurs at the dissipation region,
and creates the small magnetic islands (\cite{tan00}; \cite{tan01a}).
The islands are ejected along the current sheet.
Just after the plasmoid ejection ($t\sim 62$),
the drift velocity increases above the threshold of anomalous resistivity.
The anomalous resistivity sets in, and Petschek-like reconnection starts.
The gas is accelerated horizontally to $\sim 8-13$ (Fig. \ref{vx}{\it b}),
which is equal to the local Alfv\'en velocity.
The gas is heated to $T_{\rm max}\sim 125$
(Figs. \ref{te}{\it b} and \ref{te}{\it c}).
It is explained by
\begin{equation}
T_{\rm max}\sim\left(1+{1\over\beta}\right)
               {n_{\rm in}\over n_{\rm out}}T_{\rm in},
\label{Tmax}
\end{equation}
where $n_{\rm in}$, $n_{\rm out}$, and $T_{\rm in}$
are the densities of inflowing and outflowing gas,
and the temperature of inflowing gas.
It is derived from the equation
$n_{\rm out}kT_{\rm max}\sim n_{\rm in}kT_{\rm in}+B^2/8\pi$.
We assume $\beta\sim\beta_0\sim 0.2$,
$n_{\rm in}\sim n_{\rm out}$, and $T_{\rm in}\sim T_c\sim 25$.

Figures \ref{sshock} and \ref{fshock} show 
the slow and fast shock regions, respectively.
Two peaks in the profile of current density are
a characteristic pattern of the slow shock (see also Fig. \ref{j}c).
The fast shock is also created by the collision between 
the reconnection jet and ambient high-pressure gas (Fig. \ref{fshock}).

The magnetic energy is released at the rate of $-dE_{\rm mag}/dt\sim 35$
(Fig. \ref{energy}).
It is determined by the Poynting flux entering into the reconnection region,
\begin{equation}
-{dE_{\rm mag}\over dt}
\sim 2{B^2\over 4\pi}Sv_{\rm in}
\sim {B^3 S\over 4\pi^{3/2}\rho^{1/2} }
\end{equation}
where $v_{\rm in}$($=\epsilon v_{\rm A}$), $\epsilon$, and $S$ are
the inflow velocity, reconnection rate,
and reconnection region size, respectively.

\subsection{The Parameter Survey}

\subsubsection{The Dependence on the Existence and Direction
of the Magnetic Field in the Galactic Halo}

%Figure \ref{corona} shows 
We examine the effect of the magnetic field in the Galactic halo,
by comparing the anti-parallel-field model (typical model; model A1)
with the no-magnetic-field model (model B),
parallel-magnetic-field model (model C),
and model D (anti-parallel magnetic field in a Galactic halo
and parallel magnetic field in the other Galactic halo)
(Table \ref{parameter}).
In models B and C, no magnetic reconnection occurs.
In model C, the parallel magnetic field in the Galactic halo suppresses
Parker instability. %(Fig. \ref{energy_corona}).
The velocity is $\sim 3-6$,
which is larger than that of model B (Fig. \ref{vx}),
because the gas is compressed between 
the magnetic loop and ambient magnetic field.
In model D, the magnetic energy release rate is very small.
On the other hand, the distribution of density do not depend on 
the magnetic reconnection.

\subsubsection{The Dependence on the Magnetic Field Strength}

We examine the dependence of results on 
$B_{x0}$[$=(8\pi p_{g0}/\beta_0)^{1/2}$], i.e., $\beta_0$
(models A2 and F1-9; Table \ref{parameter}).
Figure \ref{beta}{\it a} shows the $\beta_0$-dependence of 
magnetic energy release rate.
It is determined by the Poynting flux entering the reconnection region,
\begin{eqnarray}
-{dE_{\rm mag}\over dt} &\sim& 2{B^2\over 4\pi}Sv_{\rm in}\\
&\propto& \beta_0^{-3/2},
\label{dedt}
\end{eqnarray}
where $S$ is the reconnection region size,
$v_{\rm in}$ [$=\epsilon v_{\rm A}=\epsilon B/(4\pi\rho)^{1/2}$]
is the inflow velocity to the reconnection region,
and $\epsilon$ is the reconnection rate.
It is also shown ($\sim 4.5\beta_0^{-3/2}$),
by assuming $\epsilon\sim 0.1$, $S\sim 20$, and $p_g\sim 0.6$.
It explains the results well.
Figure \ref{beta}{\it b} shows the time
when maximum heating rate is attained.
It is determined by the time scale of Parker instability,
i.e., Alfv\'en time
($\tau_{\rm A}\propto v_{\rm A}^{-1}\propto\beta_0^{1/2}$),
as shown ($\sim 100v_{\rm A}^{-1}\sim 90\beta_0^{1/2}$),
where we assume that the vertical of loop is $\sim 0.1v_{\rm A}$ 
(\cite{mat98})
and the reconnection occurs at $z\sim 10$.

Figure \ref{beta}{\it c} shows the maximum temperature ($T_{\rm max}$).
The gas is heated to 
\begin{eqnarray}
T_{\rm max}
&\sim&\left(1+{1\over\beta_0}\right)
      {n_{\rm in}\over n_{\rm out}}T_{\rm in}\\
&\propto& 1+{1\over\beta_0},
\label{Tbeta}
\end{eqnarray}
(see eq.\ [\ref{Tmax}]),
as also plotted ($\sim 25[1+\beta_0^{-1}]$),
by assuming $n_{\rm in}\sim n_{\rm out}$ and $T_{\rm in}\sim T_c\sim 25$.
It explains the results well.
Figure \ref{beta}{\it d} shows the maximum velocity.
The gas is accelerated to Alfv\'en velocity,
\begin{eqnarray}
v_{\rm jet}\sim v_{\rm A}&=&{B\over (4\pi\rho)^{1/2}}\\
&\propto& \beta_0^{-1/2},
\label{va_beta}
\end {eqnarray}
as shown by assuming $T\sim 25$.
It also explains the results well.

\subsubsection{The Dependence on the Magnetic Field in the Galactic Halo}

We examine the position ($z_h$) of anti-parallel magnetic field 
(models A2 and G1-3; Table \ref{parameter}).
In the near-magnetic-field model (model G3),
the tearing instability occurs between the Galactic disk and halo.
The magnetic field dissipates in an early phase ($t\sim 0-50$),
and the wavelength of the magnetic loops is shorter 
than that of the other models.

We also examine the existence of magnetic field in the Galactic halo,
for the near-magnetic-field models
(models G3, K, and L; Table \ref{parameter}).
In models K (the no-magnetic-field model) and G3, 
the magnetic energy is released gradually
by the tearing instability or magnetic dissipation
in an early phase ($t\sim 0-35$)
(Figs. \ref{energy_low}{\it a} and \ref{energy_low}{\it b}).
In model L (the parallel-magnetic-field model),
the dissipation is suppressed (Fig. \ref{energy_low}{\it c}).

We, furthermore, examine $\beta_0$ in the near-magnetic-field models
(models G3 and J1-8; Table \ref{parameter}),
as also shown in Figure \ref{beta} by $\triangle$.
The $\beta_0$-dependence of these models
are equal to that of high-magnetic-field models.

\subsubsection{The Dependence on the Resistivity Model\label{secres}}

We examine the resistivity model.
Figure \ref{res1} shows the reconnection rate, in the anomalous (model A1)
and the uniform resistivity model (models M and N).
In the uniform resistivity model, Sweet-Parker reconnection occurs,
so that the reconnection rate is much smaller than 
that in the anomalous resistivity model.
The reconnection rate increases transiently at $t\sim 40$.
It is different from Yokoyama \& Shibata (1996)'s results:
their reconnection rate is not time-dependent.
The reason is as follows;
In our model, the position of magnetic field in the halo is lower than theirs,
so that the magnetic field is stronger and
the current density is larger at current sheet.
The reconnection rate, however, stays at nearly steady small value,
which is consistent with the Sweet-Parker scaling: 
$\eta |J|\propto$Re$_m^{-1/2}\propto\eta^{1/2}$.

Table \ref{resmodel} shows the route to fast reconnection.
In the anomalous resistivity model, 
Petschek type reconnection occurs after tearing instability and
onsets of anomalous resistivity,
while Sweet-Parker type reconnection occurs
in the uniform resistivity model.
Figure \ref{res2} displays the dependence on the resistivity models.
It shows a characteristic patterns of two reconnection models.

\section{DISCUSSION}

\subsection{The Origin of X-Ray Gas in the Galactic Halo\label{dis2}}

Parker instability is initiated by a small perturbation,
a supernova explosion (\cite{kam97}; \cite{ste01}),
collision of the high-velocity clouds (HVCs; \cite{san00}), 
and cosmic rays etc.  
The X-ray gas can be generated,
if the magnetic reconnection is triggered by Parker instability
or the collision of HVCs (Kerp et al.\ 1994, 1996).

The reconnection heats the gas to 
\begin{equation}
T\sim 10^6 \left({n\over 10^{-3}\ {\rm cm}^{-3}}\right)^{-1}
           \left({B\over 3\ \mu{\rm G}}\right)^2\ \rm K.
\end{equation}
The reconnection also accelerates the gas to
\begin{equation}
v_{\rm A}
\sim 300 \left({n\over 10^{-3}\ {\rm cm}^{-3}}\right)^{-1/2}
         \left({B\over 3\ \mu{\rm G}}\right)\ \rm km\ s^{-1}.
\end{equation}
The duration of fast reconnection is 
\begin{eqnarray}
t&\sim& {l\over \epsilon v_{\rm A}}\\
&\sim& 3000
    \left({l\over 100\ {\rm pc}}\right)
    \left({\epsilon\over 0.1}\right)^{-1}\nonumber\\
& & \left({n\over 10^{-3}\ {\rm cm}^{-3}}\right)^{1/2}
    \left({B\over 3\ \mu{\rm G}}\right)^{-1}\ \rm yr,
\end{eqnarray}
where the reconnection rate is $\epsilon$($=v_{\rm in}/v_{\rm A}$).

The anomalous resistivity sets in 
at least when the current sheet thickness becomes 
comparable with ion Lamor radius
[$\sim 3\times 10^7$ $(T/10^4\ {\rm K})^{1/2}$ $(B/3\mu\ {\rm G})^{-1}$ cm].
The radius of magnetic island (i.e., plasmoid) is larger than it.
The island in 2D simulation is helically twisted magnetic tube in 3D.
Its volume, mass, and luminosity are
\begin{eqnarray}
V_{tube}&\sim& 10^{61}
    \left({r_p\over 10\ {\rm pc}}\right)^2
    \left({l_p\over 100\ {\rm pc}}\right)\ \rm cm^3,\\
M_{tube}&\sim& 2\times 10^{33}
    \left({r_p\over 10\ {\rm pc}}\right)^2
    \left({l_p\over 100\ {\rm pc}}\right)
    \left({n  \over 1\ {\rm cm}^{-3}}\right)\ \rm g\\
L_{tube}&\sim& n^2\Lambda (T)V_{tube}\nonumber\\
&\sim& 10^{34}
    \left({n\over 10^{-3}\ {\rm cm^{-3}}}\right)^2
    \left({\Lambda(T)\over 10^{-21}\ {\rm erg\ cm^3\ s^{-1}}}\right)\nonumber\\
& & \left({r_p\over 10\ {\rm pc}}\right)^2
    \left({l_p\over 100\ {\rm pc}}\right),
\end{eqnarray}
where $r_p$ and $l_p$ are radius and length of the plasmoid,
and $\Lambda(T)$ is the cooling function.
Then, the number of magnetic tubes at such size is $\sim 10^{7-8}$.

Total released magnetic energy is 
\begin{equation}
E_{\rm mag}\sim 10^{51} \left({B\over 3\ \mu{\rm G}}\right)^2 
\left({\lambda_{\rm tot}\over 1\ {\rm kpc}}\right)^2
\left({l\over 100\ {\rm pc}}\right)\ \rm erg,
\end{equation}
where $\lambda_{\rm tot}^2$ is the total area of many reconnection regions,
and $l$ are the typical thickness of magnetic loop.
The energy release rate, then, is
\begin{eqnarray}
{{\rm d}E_{\rm mag}\over {\rm d}t} &\sim& 10^{40}
    \left({B\over 3\ \mu{\rm G}}\right)^3 \nonumber\\
& & \left({\lambda_{\rm tot}\over 1\ {\rm kpc}}\right)^2 
    \left({n\over 10^{-3}\ {\rm cm}^{-3}}\right)^{-1/2}\ \rm erg\ s^{-1}.
\end{eqnarray}
It is also derived from eq.\ (\ref{dedt}).
It can explain the X-ray luminosity 
($\sim 10^{39-40}$ erg s$^{-1}$; \cite{pie98}).
The heated gas is confined for  
$\tau_{\rm cond}
\sim 10^9 (n/10^{-3}\ {\rm cm}^{-3})
(\lambda_{\rm eff}/3\ {\rm kpc})^2 (T/10^6\ {\rm K})^{-5/2}$ yr
(eq.\ [\ref{cond2}]),
where $\lambda_{\rm eff}$ is the effective length of
the helical magnetic tubes.
The reconnection creates the bipolar jet, 
forming the high velocity gas at Alfv\'en velocity 
($\sim 300$ km s$^{-1}$; see also \cite{nit01}, \cite{nit02}).
The Doppler shift of the bipolar jet will be the evidence
of the Galactic flare, proposed in this paper.

The plasmoid of cool gas is also created at the same time
by the reconnection (\cite{yok96}).
It is also confined by magnetic field.
The high velocity cool gas as well as hot gas will be 
the evidence of our model.

\subsection{The Comparison With Other Numerical Simulations\label{dis1}}

The 2D model examined in this paper is an extension from 
the supernova-shock driven reconnection model in the Galactic disk
(Tanuma et al.\ 1999a, 1999b, 2001a; Tanuma 2000).
Some 2D numerical simulations were done for the reconnection in the Galaxy 
(\cite{zim97}; \cite{bir98}; Tanuma et al.\ 1999a, 2001a).
They assume the current sheet for the initial condition
(see also \cite{uga92}; \cite{uga01}).
Different from their models,
Parker instability creates the current sheets spontaneously 
in our present model.

In our results, the tearing instability triggers Petschek reconnection.
Sweet-Parker current sheet, however, would be created,
and the secondary tearing instability will occur in the current sheet
before Petschek reconnection,
if we use fine grids like Tanuma et al.\ (1999a, 2001a)s' model
(Table \ref{resmodel}).
Furthermore, in the actual Galaxy ($>10^{15}$),
as well as in solar corona ($\sim 10^{12}$),
the magnetic Reynolds number is
much larger than that of the numerical simulations ($\sim 10^{4-5}$),
so that the current-sheet thickness must become
much smaller than the current-sheet length,
to set in the anomalous resistivity
(e.g., through ``fractal tearing instability''; 
\cite{tan00}; \cite{shi01}; \cite{tan01a}).

A different situation from the solar flare model is 
the growth of odd-mode of Parker instability.
Another different result from Yokoyama \& Shibata (1996)'s model is
the time variation of reconnection rate in the uniform resistivity model
(studied in section \ref{secres}).
On the other hand, in our model,
Petschek reconnection occurs in the anomalous resistivity model,
while Sweet-Parker reconnection occurs in the uniform resistivity model.
This result is the same with Yokoyama \& Shibata (1994).
We confirmed this in more general situation.

\subsection{Turbulence and 3D Effects}

The MHD turbulence may be also important in reconnection problem (\cite{laz99}).
The effective reconnection rate increases if the diffusion region
is in a state of MHD turbulence.
It is, however, difficult to resolve MHD turbulence (even if in 2D).

Recently, we revealed that 
the fast reconnection occurs after the current sheet thinning
by the tearing instability in both 2D (\cite{tan99b}) and 
3D models (\cite{tan00}; \cite{tan01b}) under a low spatial resolution
and the assumption of initial current sheet.
We found no difference between 2D and 3D models in reconnection rate, 
inflow velocity, velocity of reconnection jet, temperature of heated gas, 
slow shock formation accompanied with Petschek reconnection, 
fast shock formation due to the collision
between the reconnection jet and high pressure gas,
and time scale of these phenomena.
These results do not have a large quantitative difference from 
2D models with a high spatial resolution examined by Tanuma et al.\ (2001a)
(see also \cite{uga96}). 
Rayleigh-Taylor instability is, however, excited due to the collision,
which occurs much after the onset of anomalous resistivity 
(\cite{tan02}; in prep.\ for study of details).

We can not examine 3D model with the same small grid with that of 2D model,
so that we examine 2D model with a small grid in this paper.
The secondary tearing instability, however, may be different between 2D and 3D,
if we assume enough small grid.
It is also important to study details of 3D structure of diffusion region,
the tearing instability, plasmoid ejection, 
heating of X-ray gas in the Galactic halo,
and creation of high velocity ``bipolar jet'',
by performing 3D simulations with enough fine grid.
It is our future work.

\section{SUMMARY}

We propose the Galactic flare model for the origin
of the X-ray gas in the Galactic halo.
We examine the magnetic reconnection triggered 
by Parker instability in the Galaxy,
by performing the two-dimensional resistive MHD numerical simulations.

At the initial condition,
we assume the horizontal flux sheet in the Galactic disk,
and the anti-parallel magnetic field in the Galactic halo.
The magnetic field inflates toward the Galactic halo,
by Parker instability.
It collides with the anti-parallel magnetic field in the Galactic halo,
and creates the current sheets.
The tearing instability occurs in the current sheet,
and creates magnetic islands.
Just after the plasmoid ejection, the anomalous resistivity sets in,
and Petschek reconnection occurs.
In the Galactic halo,
the magnetic field of $B\sim$ several $\mu$G can heat the gas 
to $T\sim 10^6$ K.
If the bipolar jet of high-velocity hot gas or cool gas
at Alfv\'en velocity ($\sim 300$ km s$^{-1}$) is observed,
it is the evidence of the magnetic reconnection model in the Galactic halo.

\acknowledgments

The authors thank R. Matsumoto in Chiba University
and K. Makishima in Tokyo University for fruitful discussions.
The authors also gratefully acknowledge the constructive advice 
and useful comments of our referee toward improving our manuscript.
The numerical computations were carried out on VPP5000
at the Astronomical Data Analysis Center
of the National Astronomical Observatory, Japan,
which is an inter-university research institute of astronomy
operated by Ministry of Education, Culture, Sports, Science and Technology.
This work was partially supported by Japan Science and Technology
Cooperation (ACT-JST).

\clearpage

\scriptsize

\begin{center}
\begin{deluxetable}{lcccccccrc}
\tablecaption{Parameters. The typical model is named A1.
The unit of length is $\sim 100$ pc.
\label{parameter}}
\tablehead{
\colhead{model} &
\colhead{$\beta_0$\tablenotemark{a}} & 
\colhead{$z_h$\tablenotemark{b}} & 
\colhead{$z_c$\tablenotemark{c}} & 
\colhead{$B_u$\tablenotemark{d}} & 
\colhead{$B_l$\tablenotemark{e}} &
\colhead{$\triangle x$\tablenotemark{f} $\times$ $\triangle z$\tablenotemark{g}} &
\colhead{$N_x$\tablenotemark{h} $\times$ $N_z$\tablenotemark{i}} & 
\colhead{$L_x$\tablenotemark{j} $\times$ $L_z$\tablenotemark{k}} &
\colhead{$\eta$\tablenotemark{l}}
}
\startdata
A1& 0.200& 5.0& 6.0& $\uparrow\downarrow$&$\uparrow\downarrow$&
    $0.300\times 0.075$ & $403\times 603$ & $120\times 69.2$ &anomalous\nl 
A2& 0.200& 5.0& 6.0& $\uparrow\downarrow$&$\uparrow\downarrow$&
    $0.350\times 0.200$ & $203\times 303$ &  $70\times 60.0$ &anomalous\nl
B & 0.200& ---& 6.0& none                 &none               &
    $0.300\times 0.075$ & $403\times 603$ & $120\times 69.2$ &anomalous\nl
C & 0.200& 5.0& 6.0& $\uparrow\uparrow$  &$\uparrow\uparrow$  & 
    $0.300\times 0.075$ & $403\times 603$ & $120\times 69.2$ &anomalous\nl
D & 0.200& 5.0& 6.0& $\uparrow\downarrow$&$\uparrow\uparrow$  &
    $0.300\times 0.075$ & $403\times 603$ & $120\times 69.2$ &anomalous\nl
F1& 0.100& 5.0& 6.0& $\uparrow\downarrow$&$\uparrow\downarrow$&
    $0.350\times 0.200$ & $203\times 303$ &  $70\times 60.0$ &anomalous\nl
F2& 0.175& 5.0& 6.0& $\uparrow\downarrow$&$\uparrow\downarrow$&
    $0.350\times 0.200$ & $203\times 303$ &  $70\times 60.0$ &anomalous\nl
F3& 0.225& 5.0& 6.0& $\uparrow\downarrow$&$\uparrow\downarrow$&
    $0.350\times 0.200$ & $203\times 303$ &  $70\times 60.0$ &anomalous\nl
F4& 0.250& 5.0& 6.0& $\uparrow\downarrow$&$\uparrow\downarrow$&
    $0.350\times 0.200$ & $203\times 303$ &  $70\times 60.0$ &anomalous\nl
F5& 0.275& 5.0& 6.0& $\uparrow\downarrow$&$\uparrow\downarrow$&
    $0.350\times 0.200$ & $203\times 303$ &  $70\times 60.0$ &anomalous\nl
F6& 0.300& 5.0& 6.0& $\uparrow\downarrow$&$\uparrow\downarrow$&
    $0.350\times 0.200$ & $203\times 303$ &  $70\times 60.0$ &anomalous\nl
F7& 0.500& 5.0& 6.0& $\uparrow\downarrow$&$\uparrow\downarrow$&
    $0.350\times 0.200$ & $203\times 303$ &  $70\times 60.0$ &anomalous\nl
F8& 1.000& 5.0& 6.0& $\uparrow\downarrow$&$\uparrow\downarrow$&
    $0.350\times 0.200$ & $203\times 303$ &  $70\times 60.0$ &anomalous\nl
F9& 1.200& 5.0& 6.0& $\uparrow\downarrow$&$\uparrow\downarrow$&
    $0.350\times 0.200$ & $203\times 303$ &  $70\times 60.0$ &anomalous\nl
G1& 0.200& 6.0& 7.0& $\uparrow\downarrow$&$\uparrow\downarrow$&
    $0.350\times 0.200$ & $203\times 303$ &  $70\times 60.0$ &anomalous\nl
G2& 0.200& 4.5& 5.5& $\uparrow\downarrow$&$\uparrow\downarrow$&
    $0.350\times 0.200$ & $203\times 303$ &  $70\times 60.0$ &anomalous\nl
G3& 0.200& 4.0& 5.0& $\uparrow\downarrow$&$\uparrow\downarrow$&
    $0.350\times 0.200$ & $203\times 303$ &  $70\times 60.0$ &anomalous\nl
J1& 0.175& 4.0& 5.0& $\uparrow\downarrow$&$\uparrow\downarrow$&
    $0.350\times 0.200$ & $203\times 303$ &  $70\times 60.0$ &anomalous\nl
J2& 0.180& 4.0& 5.0& $\uparrow\downarrow$&$\uparrow\downarrow$&
    $0.350\times 0.200$ & $203\times 303$ &  $70\times 60.0$ &anomalous\nl
J3& 0.225& 4.0& 5.0& $\uparrow\downarrow$&$\uparrow\downarrow$&
    $0.350\times 0.200$ & $203\times 303$ &  $70\times 60.0$ &anomalous\nl
J4& 0.250& 4.0& 5.0& $\uparrow\downarrow$&$\uparrow\downarrow$&
    $0.350\times 0.200$ & $203\times 303$ &  $70\times 60.0$ &anomalous\nl
J5& 0.275& 4.0& 5.0& $\uparrow\downarrow$&$\uparrow\downarrow$&
    $0.350\times 0.200$ & $203\times 303$ &  $70\times 60.0$ &anomalous\nl
J6& 0.300& 4.0& 5.0& $\uparrow\downarrow$&$\uparrow\downarrow$&
    $0.350\times 0.200$ & $203\times 303$ &  $70\times 60.0$ &anomalous\nl
J7& 0.500& 4.0& 5.0& $\uparrow\downarrow$&$\uparrow\downarrow$&
    $0.350\times 0.200$ & $203\times 303$ &  $70\times 60.0$ &anomalous\nl
J8& 1.000& 4.0& 5.0& $\uparrow\downarrow$&$\uparrow\downarrow$&
    $0.350\times 0.200$ & $203\times 303$ &  $70\times 60.0$ &anomalous\nl
K & 0.200& ---& 5.0& none                & none               &
    $0.350\times 0.200$ & $203\times 303$ &  $70\times 60.0$ &anomalous\nl
L & 0.200& 4.0& 5.0& $\uparrow\uparrow$  &$\uparrow\uparrow$  &
    $0.350\times 0.200$ & $203\times 303$ &  $70\times 60.0$ &anomalous\nl
M & 0.200& 5.0& 5.0& $\uparrow\downarrow$&$\uparrow\downarrow$&
    $0.300\times 0.075$ & $403\times 603$ & $120\times 69.2$ &uniform($\eta=0.05$)\nl
N & 0.200& 5.0& 5.0& $\uparrow\downarrow$&$\uparrow\downarrow$&
    $0.300\times 0.075$ & $403\times 603$ & $120\times 69.2$ &uniform($\eta=0.10$)\nl 
\enddata
\tablenotetext{a}{The ratio of the gas to magnetic pressure}
\tablenotetext{b}{The position of magnetic field in the Galactic halo}
\tablenotetext{c}{The position of hot gas in the Galactic halo ($z_c=z_h+1$)}
\tablenotetext{d}{The magnetic field in the upper corona
($\uparrow\uparrow$:parallel, $\uparrow\downarrow$:anti-parallel)}
\tablenotetext{e}{The magnetic field in the lower corona
($\uparrow\uparrow$:parallel, $\uparrow\downarrow$:anti-parallel)}
\tablenotetext{f}{The grid size in $x$-axis (uniform)}
\tablenotetext{g}{Minimum grid size in $z$-axis}
\tablenotetext{h}{The number of grids in $x$-axis}
\tablenotetext{i}{The number of grids in $z$-axis}
\tablenotetext{j}{The simulation region size in $x$-axis}
\tablenotetext{k}{The simulation region size in $z$-axis}
\tablenotetext{l}{The resistivity model}
\end{deluxetable}
\end{center}

\clearpage

\normalsize

\begin{center}
\begin{deluxetable}{ccccc}
\tablecaption{Variables at the initial condition 
for the typical model (model A1).
\label{normalization}}
\tablewidth{0pt}
\tablehead{
\colhead{variable} &
\colhead{cool dense disk} &
\colhead{} & 
\colhead{hot rarefied halo} &
\colhead{}
}
\tablewidth{0pt}
\startdata
$T$
& $T_0=1$    & ($|z|<z_c$)  
& $T=25T_0$  & ($|z|>z_c$) \nl
$\rho$
& $\rho_0=1$             & ($|z|=0$) 
& $\rho\sim 0.004\rho_0$ & ($|z|\sim z_h$) \nl
$p_g$
& $p_{g0}=0.6$        & ($|z|=0$)
& $p_g\sim 0.1p_{g0}$ & ($|z|\sim z_h$) \nl
$B_x$
& $B_{x0}\sim 8.68$   & ($|z|=0$)  
& $B_x\sim 0.3B_{x0}$ & ($|z|\sim z_h$) \nl
$p_{\rm tot}$ 
& $p_{\rm tot,0}=3.6$                & ($|z|=0$)
& $p_{\rm tot}\sim 0.1p_{\rm tot,0}$ & ($|z|\sim z_h$) \nl
$\beta_0$ 
& $\beta_0=0.2$ & ($|z|=0$)  
& $\beta_0=0.2$ & ($|z|>z_h$) \nl
\enddata
\end{deluxetable}
\end{center}

\clearpage

\begin{center}
\begin{deluxetable}{llll}
\tablecaption{
The route to the fast reconnection.
Parker instability trigger model is examined in this paper.
In this case, resistivity model determines the route.
The supernova shock trigger model is examined in Tanuma et al.\ (2001a).
\label{resmodel}}
\tablewidth{0pt}
\tablehead{
\colhead{} &
\colhead{Parker instability trigger} &
\colhead{Parker instability trigger} &
\colhead{supernova shock trigger}
}
\tablewidth{0pt}
\startdata
trigger& Parker instability& Parker instability& supernova shock\nl
resistivity model& anomalous& uniform       & anomalous\nl
\hline
evolution
 & downflow along rising loop& downflow along rising loop& tearing instability\nl
 & sheet thinning       & sheet thinning    & sheet thinning       \nl
 & $\downarrow$         & Sweet-Parker sheet& Sweet-Parker sheet   \nl
 & tearing instability  & --                & secondary tearing    \nl
 &further sheet thinning& --                &further sheet thinning\nl
 & anomalous resistivity& --                & anomalous resistivity\nl
 & Petschek reconnection& --                & Petschek reconnection\nl
\enddata
\end{deluxetable}
\end{center}

\clearpage

\begin{figure}[p]
\caption{Schematic illustration of our numerical simulations.
(a) As the initial condition,
we assume the horizontal, parallel magnetic field 
and cool, dense gas in the Galactic disk,
and the nearly uniform magnetic field
which is anti-parallel to the disk field
and hot, rarefied gas in the Galactic halo.
The anti-parallel field is created, for example, by Coliori's force.
(b) Parker instability is initiated by the random perturbations
in the Galactic disk.
The magnetic field in the Galactic disk bends across the equatorial plane.
(c) The inflating magnetic loop collides with 
the anti-parallel magnetic field in the Galactic halo. 
The magnetic reconnection occurs.
(d) The heated gas is confined by the magnetic field in the Galactic halo 
for a long time.
\label{picture}}

\caption{The temperature ($T$) in the unit of $T_0\sim 10^4$ K,
magnetic field lines, and velocity vectors, for a typical model (model A1),
The units of length, velocity, and time are 
1 kpc, 10 km s$^{-1}$, $10^7$ yr, respectively.
(a) Parker instability occurs in the Galactic disk.
The magnetic field of the Galactic disk inflates to the Galactic halo.
(b) The inflating magnetic loop collides with anti-parallel magnetic field
in the Galactic halo. 
The current sheets are created.
(c) The magnetic reconnection occurs at $t\sim 62$,
and heats the gas to $T_{\rm max}\sim 125$.
(d) The heated gas is confined by the magnetic field in the Galactic halo.
\label{te}}

\end{figure}
\begin{figure}[htb]

\caption{Same with Figure \ref{te}, 
but the density ($\rho$) in the unit of $10^{-25}$ g. 
(b) The gas, which is initially between the Galactic disk and halo,
are raised by the inflating magnetic loops.
(d) The large high-density clouds are created at the valleys
of magnetic loops, and will become star forming regions.
\label{de}}

\end{figure}
\begin{figure}[p]

\caption{Same with Figure \ref{te}, but the current density ($J$).
(b) Eight magnetic loops collide with anti-parallel magnetic field
in the Galactic halo, so that the current density increases there.
(d) The magnetic field confines the high-current-density regions
at the valleys of magnetic loops near the Galactic plane.
\label{j}}

\caption{Time variations of
(a) drift velocity ($v_d=|J|/\rho$) and (b) velocity ($|v_x|$),
for model A1 (solid lines; typical model, i.e., 
anti-parallel-magnetic-field model),
model B (dotted lines; the no-magnetic-field model),
model C (dashed lines; the parallel-magnetic-field model),
and model D (dashed and dotted lines; 
the anti-parallel magnetic field in a halo and 
parallel magnetic field in the other halo).
(a) In models A1 and D, the drift velocity increases steeply above 
the threshold ($v_c=400$) of anomalous resistivity at $t\sim 62-65$.
(b) In all models, the velocity increases gradually by Parker
instability to $\sim 2.5-3$ ($t\sim 40-62$). 
In models A1 and D, by the magnetic reconnection ($t>62$),
the velocity increases steeply to $\sim 8-13$,
which is equal to the local Alfv\'en velocity in the Galactic halo.
\label{vx}}

\end{figure}
\begin{figure}[htb]

%\caption{Schematic illustrations of Figure \ref{vx}.
%(a) The drift velocity increases steeply
%above the threshold of anomalous resistivity,
%when the magnetic loop collides with
%the ambient anti-parallel magnetic field in the Galactic halo.
%(b) The velocity increases steeply 
%to the local Alfv\'en velocity in the Galactic halo
%by the magnetic reconnection.
%\label{vxfig}}

\caption{Time variations of the various energies.
(a) The gravitational energy is released
by Parker instability in its nonlinear phase ($t\sim 40-62$).
The magnetic energy is released 
mainly by the magnetic reconnection ($t>62$).
(b) The magnetic energy release rate increases
to $-dE_{\rm mag}/dt\sim 75$ by the magnetic reconnection ($t\sim 70$).
\label{energy}}

\end{figure}
\begin{figure}[htb]

\caption{The profile of the variables in $x=-13.8$ at $t=70$.
The magnetic reconnection occurs around $(x, z)\sim(-26, 7.8)$.
The fast reconnection such as Petschek model is accompanied by 
two slow shock regions along the reconnection jets.
The profile of current density ($|J|$) has two peaks
at two slow shock regions.
\label{sshock}}

\caption{Same with Figure\ref{sshock}, but $z=6.5$.
The gas is accelerated, by magnetic tension force,
to the local Alfv\'en velocity in the Galactic halo,
and collides with the ambient gas.
The fast shock region is created around $x\sim -10$.
\label{fshock}}

\end{figure}
\begin{figure}[htb]

%\caption{The density ($\rho$) in the unit of $10^{-25}$ g, 
%magnetic field lines, and velocity vectors,
%for (a) model A1 (the anti-parallel-magnetic-field model),
%(b) model B (the no-magnetic-field model),
%(c) model C (the parallel-magnetic-field model),
%and (d) model D (anti-parallel magnetic field in one halo 
%and and parallel magnetic field in the other halo).
%In models B and C, no magnetic reconnection occurs.
%In model C, the ambient parallel magnetic field in the Galactic halo
%suppresses the inflation of the magnetic loops from the Galactic disk.
%\label{corona}}

%\caption{The time variation of the various energies,
%for model B (the no-magnetic-field model),
%model C (the parallel-magnetic-field model), 
%and model D (anti-parallel magnetic field in a halo 
%and parallel magnetic field in the other halo).
%In models B and C,
%about a half of the thermal energy is supplied only by Parker instability.
%In model D (and model A1), 
%the magnetic reconnection converts the magnetic energy to thermal one.
%\label{energy_corona}}

\end{figure}
\begin{figure}[htb]

\caption{The dependence of the results on $\beta_0$,
i.e., the magnetic field strength [$B_x=(8\pi p_g/\beta_0)^{1/2}$] 
(models A2, F1-0, G3, and J1-8).
The results of the far-magnetic-field models (models A2 and F1-9) 
and the near-magnetic-field models (models G3 and J1-8) are
shown by $\diamond$ (with the dashed lines) 
and $\triangle$ (with the solid lines), respectively.
(a) The maximum magnetic energy release rate ($-dE_{\rm mag}/dt$).
It is determined by Poynting flux
($\propto\beta_0^{-3/2}$; the dashed line).
(b) The time when the maximum magnetic energy release rate is attained.
The time scales of these phenomena depend on Alfv\'en time 
($\propto v_{\rm A}^{-1}\propto\beta_0^{1/2}$; the dashed lines).
(c) The maximum temperature ($T_{\rm max}$) of heated gas.
It is determined by the released magnetic energy 
($\propto 1+1/\beta_0$; the dashed line).
(d) Maximum $v_x$, which is determined by Alfv\'en velocity 
($v_{\rm A}\propto\beta_0^{-1/2}$; the dashed line).
\label{beta}}

\end{figure}
\begin{figure}[htb]

%\caption{The mass density ($\rho$), magnetic field lines, 
%and velocity vectors, for model G3 (the near-magnetic-field model).
%The magnetic field starts to dissipate 
%by the tearing instability or magnetic dissipation.
%The wavelength of the magnetic loops is shorter 
%than that of the typical model. 
%\label{low}}

\caption{Time variations of the various energies,
for the near-magnetic-field models (models G3, K, and L).
(a) In model G3 (the anti-parallel-magnetic-field model),
the magnetic field dissipates 
by the tearing instability or magnetic dissipation ($t\sim 0-40$).
The wavelength of the magnetic loops is shorter 
than that of the typical model. 
(b) In model K (the no-magnetic-field model),
the magnetic dissipation occurs in an early phase.
(c) In model L (the parallel-magnetic-field model),
the magnetic field does not dissipate in an early phase,
because the parallel field of the Galactic halo suppresses the dissipation.
\label{energy_low}}

\end{figure}
\begin{figure}[htb]

\caption{The time variation of the reconnection rate ($\eta |J|$),
in the anomalous resistivity model (model A1)
and the uniform resistivity model
($\eta=0.05$ [the dashed line; model M]
and $\eta=0.10$ [the dotted line; model N]).
Petschek reconnection occurs in the anomalous resistivity model.
On the other hand, Sweet-Parker reconnection occurs,
in the uniform resistivity model.
\label{res1}}

\caption{The current density, magnetic field lines, and velocity vectors,
for (a) the anomalous resistivity model (model A1)
and (b) the uniform resistivity model ($\eta=0.05$; model M]).
(a) Petschek reconnection occurs in the anomalous resistivity model.
The small dissipation region and slow shock regions are
the characteristics of Petschek reconnection.
(b) In the uniform resistivity model, the long current sheet forms 
because Sweet-Parker reconnection occurs.
\label{res2}}

\end{figure}
\clearpage

%\newpage
%\plotone{f1.eps}
%\newpage
%\plotone{f2.eps}
%\newpage
%\plotone{f3.eps}
%\newpage
%\plotone{f4.eps}
%\newpage
%\plotone{f5.eps}
%\newpage
%\plotone{f6.eps}
%\newpage
%\plotone{f7.eps}
%\newpage
%\plotone{f8.eps}
%\newpage
%\plotone{f9.eps}
%\newpage
%\plotone{f10.eps}
%\newpage
%\plotone{f11.eps}
%\newpage
%\plotone{f12.eps}
%\newpage

\end{document}